\documentclass[namedreferences, final]{solarphysics}

\usepackage[hyperref,optionalrh,showbiblabels]{spr-sola-addons} 
\usepackage{graphicx}        
\usepackage{color}           
\usepackage{breakurl}        
\usepackage{ulem}
\usepackage{multirow}

\newcommand{\cw}{\columnwidth}

\chardef\us=`\_

\newcommand{\cmcmsr}{~cm$^2$\,sr}
\newcommand{\intunit}{~(cm$^2$\,sr\,s\,MeV)$^{-1}$}

\newcommand{\circled}[1]{\raisebox{.5pt}{\textcircled{\raisebox{-.9pt}{#1}}}}

\newcommand{\emaxe}{E_{\mathrm{max},e}}
\newcommand{\emaxp}{E_{\mathrm{max},p}}

\begin{document} \begin{article}

\begin{opening}
\title{The Solar Electron and Proton Telescope aboard STEREO -- understanding proton spectra}

\address[id=aff1]{Christian-Albrechts-Universit\"at zu Kiel, Leibnizstr. 11, 24118 Kiel, Germany}
\author[addressref={aff1},email={wraase@physik.uni-kiel.de}]{\inits{S.}\fnm{S.}~\lnm{Wraase}}
\author[addressref={aff1},corref,email={heber@physik.uni-kiel.de}]{\inits{B.}\fnm{B.}~\lnm{Heber}}
\author[addressref={aff1}]{\inits{S.}\fnm{S.}~\lnm{B\"ottcher}}
\author[addressref={aff1}]{\inits{N.}\fnm{N.}~\lnm{Dresing}}
\author[addressref={aff1}]{\inits{P.K.}\fnm{P.}~\lnm{K\"uhl}}
\author[addressref={aff1}]{\inits{R.M$^2$}\fnm{R.}~\lnm{M\"uller-Mellin}}

\runningauthor{Wraase et al.}
\runningtitle{Understanding SEPT quiet time measurements}

\begin{abstract}
The Solar Electron and Proton Telescope (SEPT) aboard the Solar Terrestrial Relations Observatory (STEREO) is designed to provide the three-dimensional distribution of energetic electrons and protons with good energy and time resolution. Each SEPT instrument consists of two double-ended magnet/foil particle telescopes which cleanly separate and measure electrons in the energy range from 30~keV to 400~keV and protons from 60~keV to 7000~keV. Anisotropy information on a non spinning spacecraft is provided by two separate but identical instruments: SEPT-E aligned along the Parker spiral magnetic field in the ecliptic plane along looking both towards and away from the Sun, and SEPT-NS aligned vertical to the ecliptic plane looking towards North and South. The dual set-up refers to two adjacent sensor apertures for each of the four viewing directions SUN, ANTISUN, NORTH, and SOUTH: one for protons, one for electrons. In this contribution a simulation of SEPT utilizing the GEANT4 toolkit has been set up with an extended instrument model in order to calculate improved response functions of the four different telescopes. This will help to understand and correct instrumental effects in the measurements.
\end{abstract}
\end{opening}

\section{Introduction} \label{S-Introduction} 
\begin{figure}
\centering
\includegraphics[width=0.6\cw]{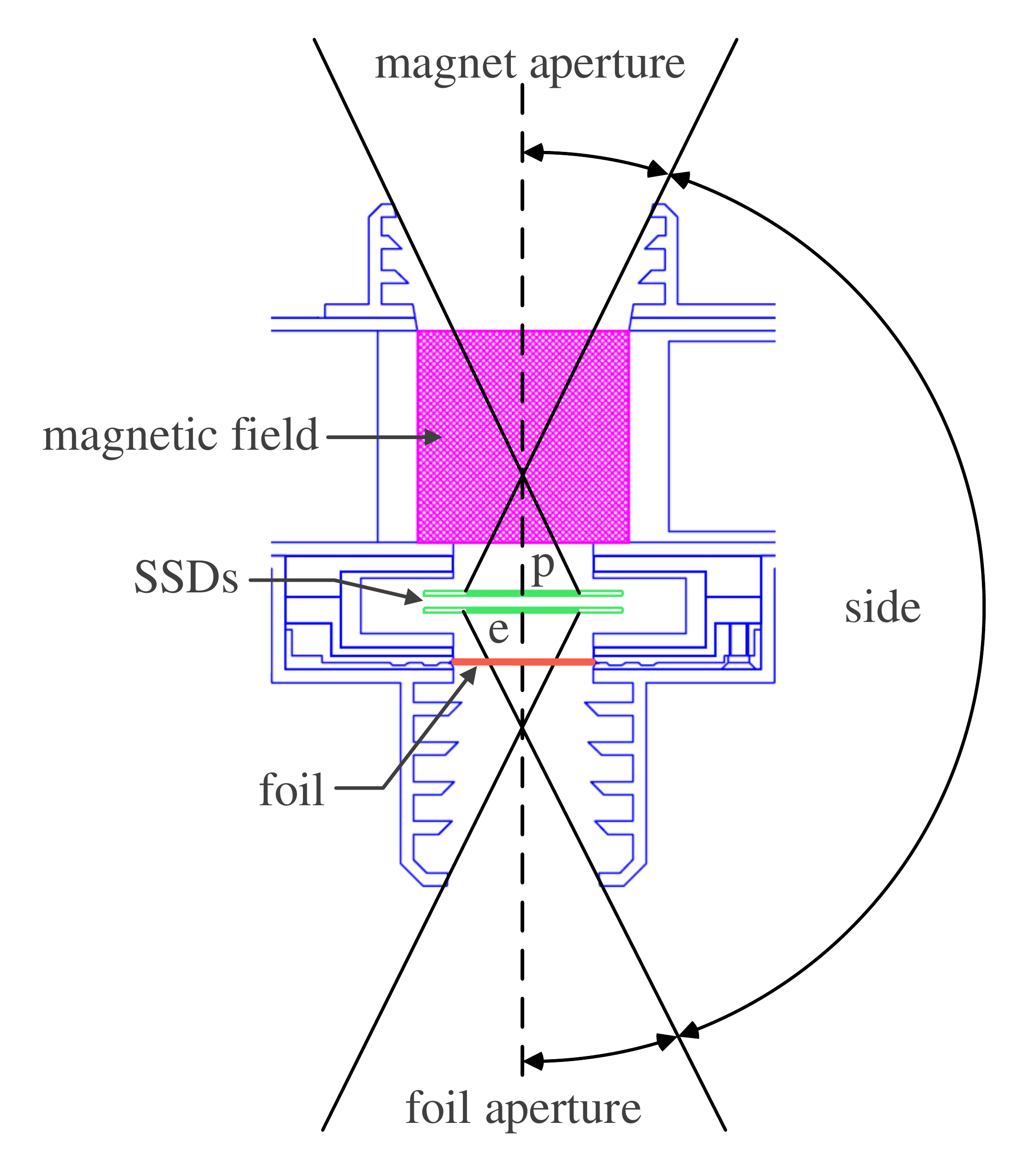}
\caption{Sketch of one of the SEPT double ended telescopes consisting of two SSD with one looking through a foil and the other one through a magnetic field. In addition to both nominal viewing directions the side direction is shown too.} 
\label{fig:sketch}
\end{figure}

Measurements of electrons in the energy range of a few ten up to several 100~keV are based on the magnet-foil technique. This technique is used by the Solar Electron and Proton Telescope \citep[SEPT,][]{Mueller-Mellin-etal-2008} aboard the two Solar Terrestrial Relations Observatory \citep[STEREO,][]{Kaiser-etal-2008} spacecraft. In order to separate electrons from ions each sensor consists of two double-ended telescopes, as sketched in Fig.~\ref{fig:sketch}, with one end of each telescope measuring primarily electrons and the other one ions. Each telescope has two 300~$\mu$m thick solid-state detectors (SSDs) that are operated in anticoincidence. While one SSD looks through an absorption foil at one end of the telescope, the other SSD looks through the gap of a magnet at the other end, hence through a magnetic field.

The foil leaves the electron spectrum essentially unchanged, but stops protons with energies up to the energy $\emaxe$. At the same energy, electrons start to penetrate the 300~$\mu$m thick SSD and will therefore trigger the anticoincidence. Only in the absence of ions with higher energies ($E>\emaxe$), the foil SSD detects electrons only. The magnetic field on the other end is designed to sweep away electrons that would not penetrate the SSD ($E<\emaxe$), but is too weak to affect ions. Therefore, in the absence of electrons with higher energies ($E>\emaxe$), the magnet SSD only detects ions. The upper energy limit of ions stopping in the magnet SSD, for which the energy spectrum can be measured quite cleanly, is determined by the energy $\emaxp$ at which ions start to penetrate the SSD. At the other end, the contribution of ions, which are able to penetrate the foil ($E>\emaxe$), to the energy spectrum of the foil SSD can then be computed and subtracted from the observed one to obtain the pure electron spectrum.

However, because electrons can scatter out of the SSD without depositing their entire energy in the detector, the measured energy can be lower than the incident energy. This effect was examined for the Electron Proton Alpha Monitor (EPAM) \citep{Gold-etal-1998}, a magnet-foil particle telescope aboard the Advanced Composition Explorer (ACE). A simplified GEANT4 \citep{Agostinelli-etal-2003} simulation model shows that the response functions of the Deflected Electron (DE) channels are not boxcars in  the nominal energy range \citep{Haggerty-etal-2003}. Further, \citet{Morgado-etal-2015} recently investigated the response of the Low Energy Magnetic Spectrometers (LEMS) and the Low Energy Foil Spectrometers (LEFS), which utilize the magnet-foil technique as well, for relativistic protons penetrating the instrument structure at oblique angles. Because of the gap between the two SSDs, particles penetrating the telescope on such trajectories can pass one detector without triggering the second one and thus trigger different proton channels. By applying the detector response for protons above several tenth of MeV the authors have investigated the May 17, 2012 solar energetic particle event and found that the data are in good agreement with results published previously by \citet{Mishev-etal-2014} and \citet[][and references therein]{GOpalswamy-etal-2013}. Here we investigate the response of the four heads of SEPT for protons that impinge on the instrument isotropically.

\begin{figure}
\centering
\includegraphics[width=0.6\cw]{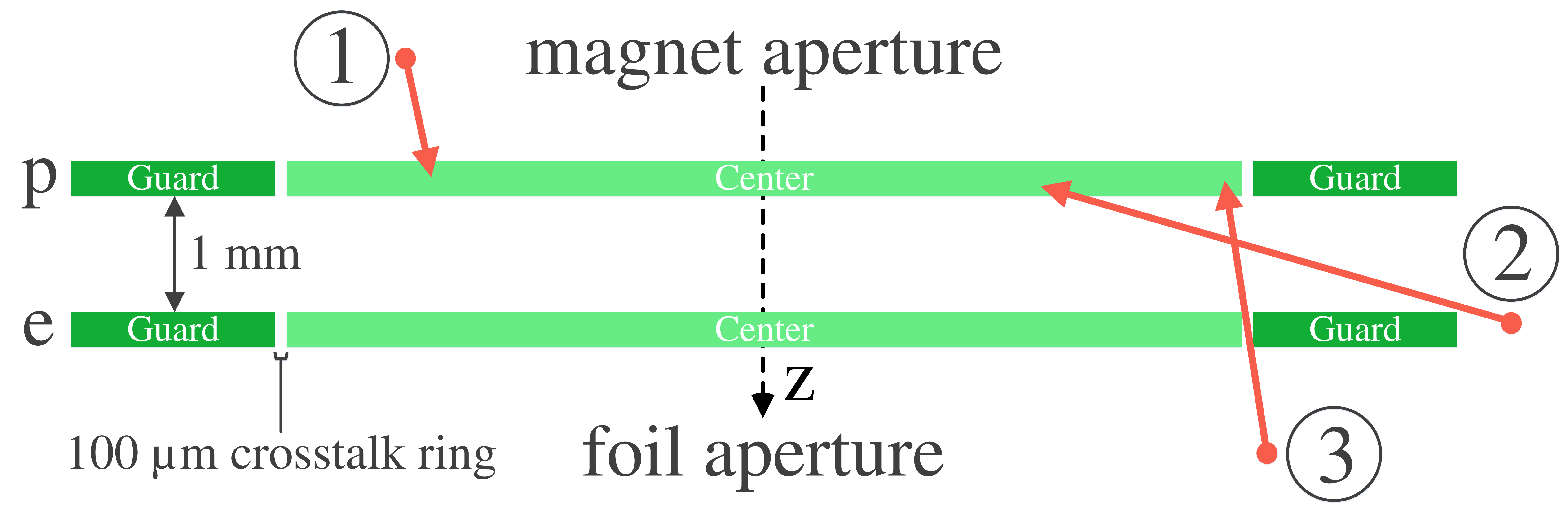}
\caption{Accurately enlarged sketch of the detector setup by two SSDs.}
\label{fig:SSD}
\end{figure}

Figure~\ref{fig:sketch} sketches the realization of the magnet foil technique by SEPT. The nominal opening angles of the ion (magnet aperture) and electron sides (foil aperture) are 52.8$^\circ$ and 52$^\circ$, respectively. All other possible viewing directions are indicated by ``side" in the figure. Since the nominal opening angle of the ion telescope is defined by the active area of the SSD facing the magnet side and the aperture of the aluminum shielding, high energy particles entering the telescope from an oblique angel add to the count rates. This effect can only be avoided if an active anticoincidence surrounding the side of the detector system and covering the oblique viewing directions, would be added. In order to minimize the probability of high energy particles passing only one detector without triggering the other one, active guard rings have been introduced to the SEPT SSDs. The corresponding two detector setup utilized by SEPT is displayed in Fig.~\ref{fig:SSD}. 

The two center elements are operated in anticoincidence with each other and with the outer guard rings. A possible trajectory of a particle entering the detector system through the apertures and hence in a nominal viewing direction is indicated by \circled{1} in Fig.~\ref{fig:SSD}. Due to engineering issues the two SSDs of each telescope are separated by 1~mm allowing oblique trajectories (track \circled{2} in Fig.~\ref{fig:SSD}) that hit one SSD without triggering any other active element and therefore without triggering the anticoincidence. Moreover, a so called crosstalk ring with a width of 100~$\mu$m separates the center and guard segments in order to decouple both from each other. Particles passing this small volume are neither counted in the center element nor in the guard ring of the corresponding SSD and can therefore contribute to the count rates of the electron and ion telescopes. A potential track is indicated by \circled{3} in Fig.~\ref{fig:SSD}. We note that particles on tracks \circled{2} and \circled{3} do not necessarily deposit their entire incident energy in the center elements and can penetrate the SSDs without triggering the anticoincidence. In what follows we will calculate the instrument response function of SEPT for isotropic proton fluxes from 20~keV up to 20~GeV utilizing the GEANT4 toolkit.

\section{GEANT4 Modeling of SEPT} \label{S-GEANT4}      
In \citet{Mueller-Mellin-etal-2008} geometry factors for both the electron and proton detectors have been calculated using GEANT4 and a simplified instrument model that represents just one of the double-sided SEPT telescopes. This was feasible because particles entering the detector system outside of the nominal view, hence particles with tracks indicated by \circled{2} in Fig.~\ref{fig:SSD}, were neglected in the previous simulation. Thus for extending the SEPT response function to higher energies and to an isotropic incidence of particles the following adaptations to the instrument model had to be made.

\subsection{SEPT and Satellite Structure}
\begin{figure}
\centering
\includegraphics[width=0.9\cw]{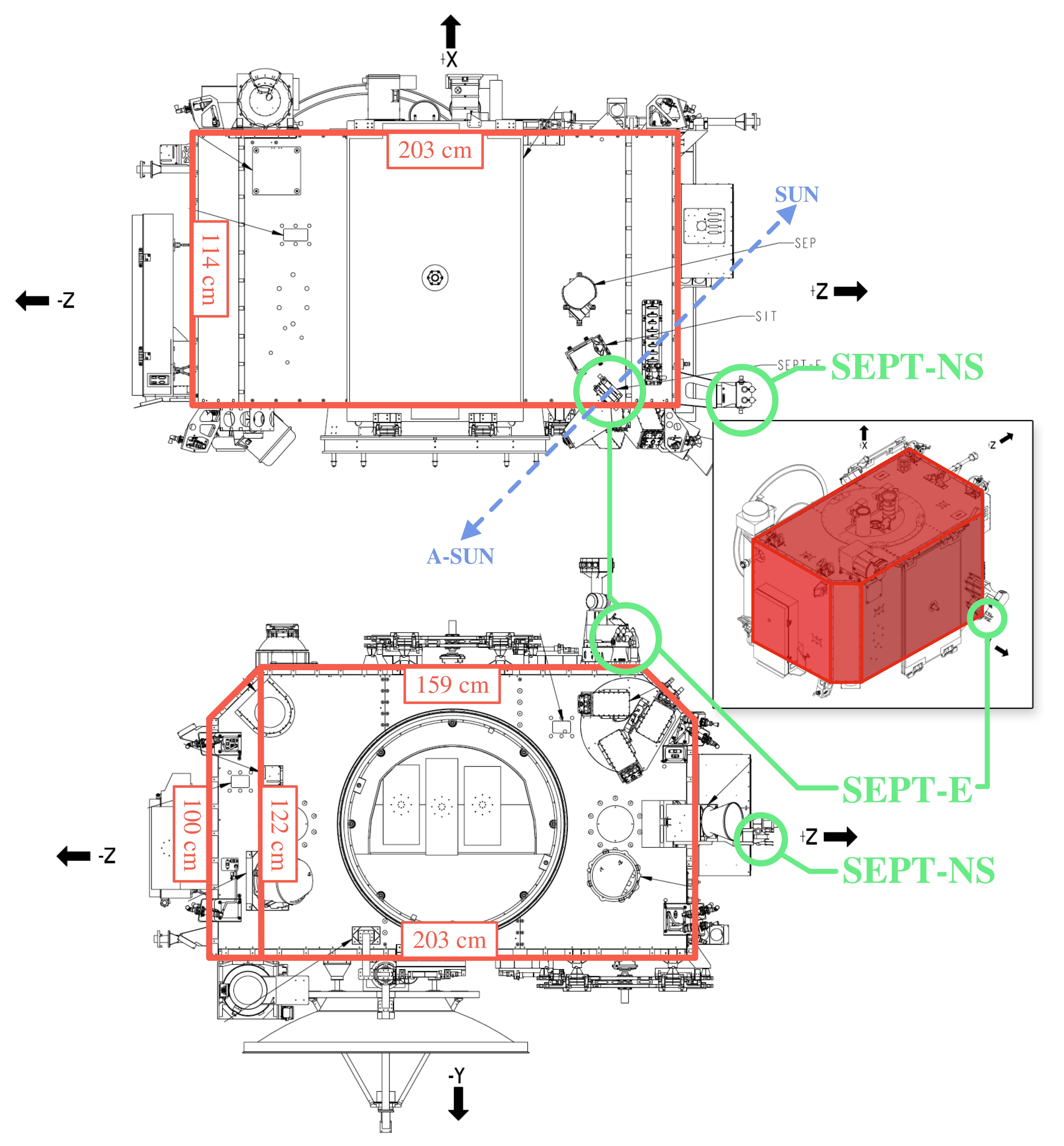}
\caption{Locations of the two SEPT instruments on the STEREO-A satellite. The shape of the red box is used to simulate the spacecraft.}
\label{fig:sept-on-stereo-a-dims}
\end{figure}

\begin{figure}
\centering
\includegraphics[width=0.6\cw]{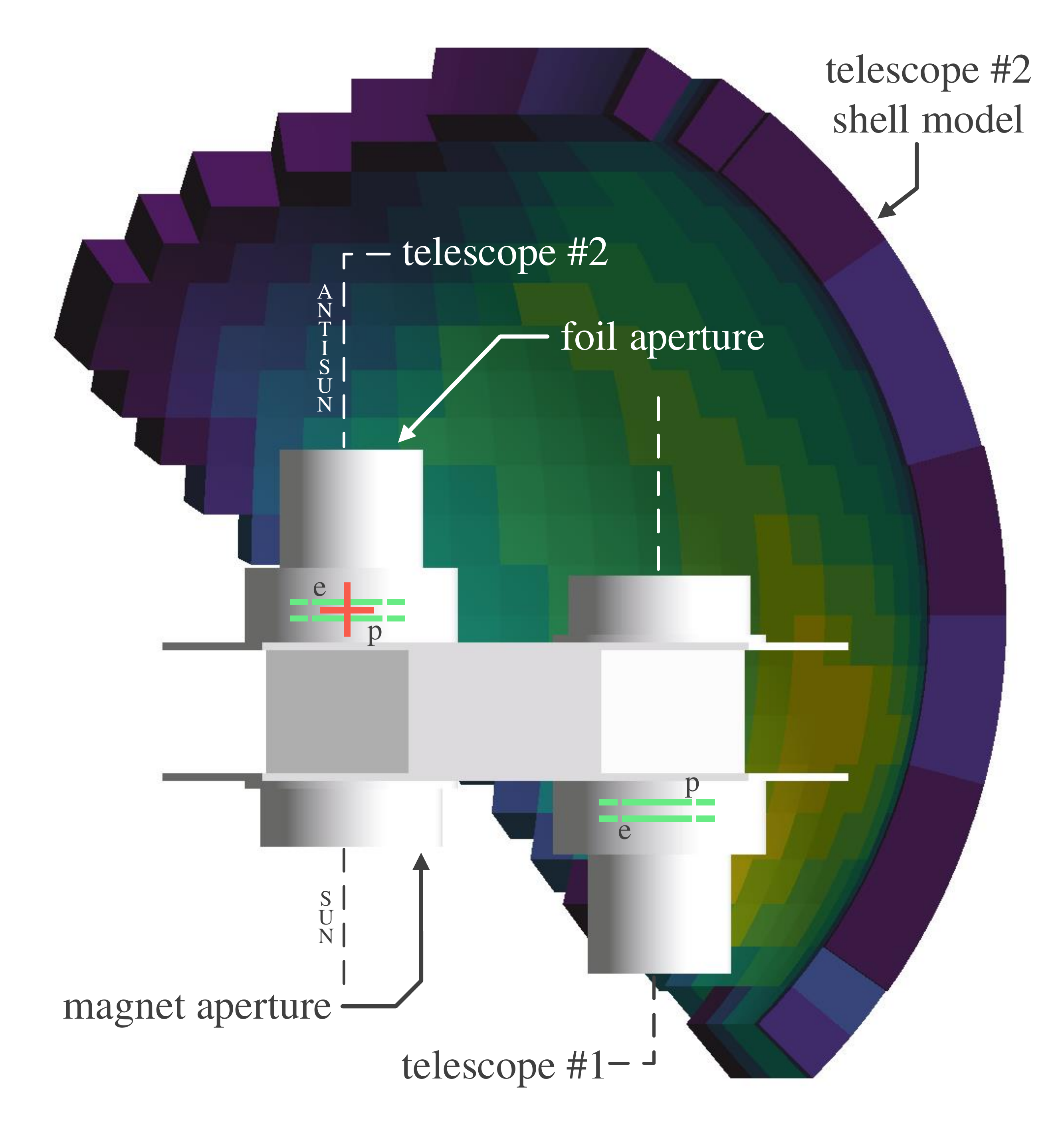}
\caption{Simulation model for telescope \#2 of SEPT-E on STA using a spherical shell to approximate the shielding effect of the s/c. The density of each shell section is roughly sketched from low to high using a color gradient from blue to yellow, respectively. The red cross indicates the center of the shell at the central point between the SSDs of telescope \#2. This model is used to simulate telescope \#2 only. An individual shell was created for telescope \#1.}
\label{fig:sept-e-sta-sphere-model}
\end{figure}

In a first step the simplified GEANT4 model of the SEPT was extended to represent the full dual telescope setup of the instrument instead of just one double-ended telescope. Additionally a representation of the passive material directly surrounding the telescopes at their sides was added to the model. Because the field of view for particles entering the detector systems from the side is partially affected by the STEREO spacecraft (s/c), efforts were made to include the s/c in the simulation. The mounting positions of the two SEPT instruments on STEREO-A (STA) are shown in Fig.~\ref{fig:sept-on-stereo-a-dims} marked by the green circles. The dimensions of the red box in the figure give the structure of the spacecraft as it was included in the simulation model (further called ``box model"). Since a more detailed model of the spacecraft is not available to us the spacecraft was assumed to be made out of aluminum with the dimensions that are shown in Fig.~\ref{fig:sept-on-stereo-a-dims} resulting in a volume of $2.77$~m$^3$ with the total mass of $620$~kg equally distributed over the whole s/c leading to a density of $\rho = 0.224$~g/cm$^3$. A further approximation was introduced due to the huge difference in size between the s/c and the telescope heads. For an isotropic simulation this difference lead to either unfeasibly long computation times or, when limiting the amount of simulated particles to a feasible level, to insufficient statistics. Therefore we chose a simplified approach: 

We surrounded the SEPT model by a spherical shell of constant width and an inner and outer radius of 75~mm and 85~mm, respectively, as displayed in Fig. \ref{fig:sept-e-sta-sphere-model}. Each shell segment reflects the mean mass and consequently the mean density of the s/c that non interacting particles have to pass before they reach the mid point between the two SSDs of one telescope. We note that therefore, the mass distribution of the s/c box model is correctly reflected only at the center of the shell at this mid point. Because the detector system of each telescope has a different mid point, an individual shell structure was computed for each telescope. For these calculations GEANT4 simulations with non interacting particles called Geantinos were performed. Utilizing this approach we approximate the influence of the s/c mass on the mean energy loss by ionization for ions. Figure \ref{fig:sept-e-sta-sphere-model} shows the resulting simulation model (further called ``shell model") for telescope \#2 of the SEPT-E instrument on STA, which reflects to the SUN and ANTISUN directions for ions and electrons, respectively (see Fig. \ref{fig:sept-on-stereo-a-dims}).

In order to validate our approximation we compared the shielding effect of the shell model with computations utilizing the box model. Since the computation time when using the box model increases by a factor of roughly 400 compared to the shell model, only a simulation with significantly lower statistics was performed.
\begin{figure}
\centering
\includegraphics[width=0.99\cw]{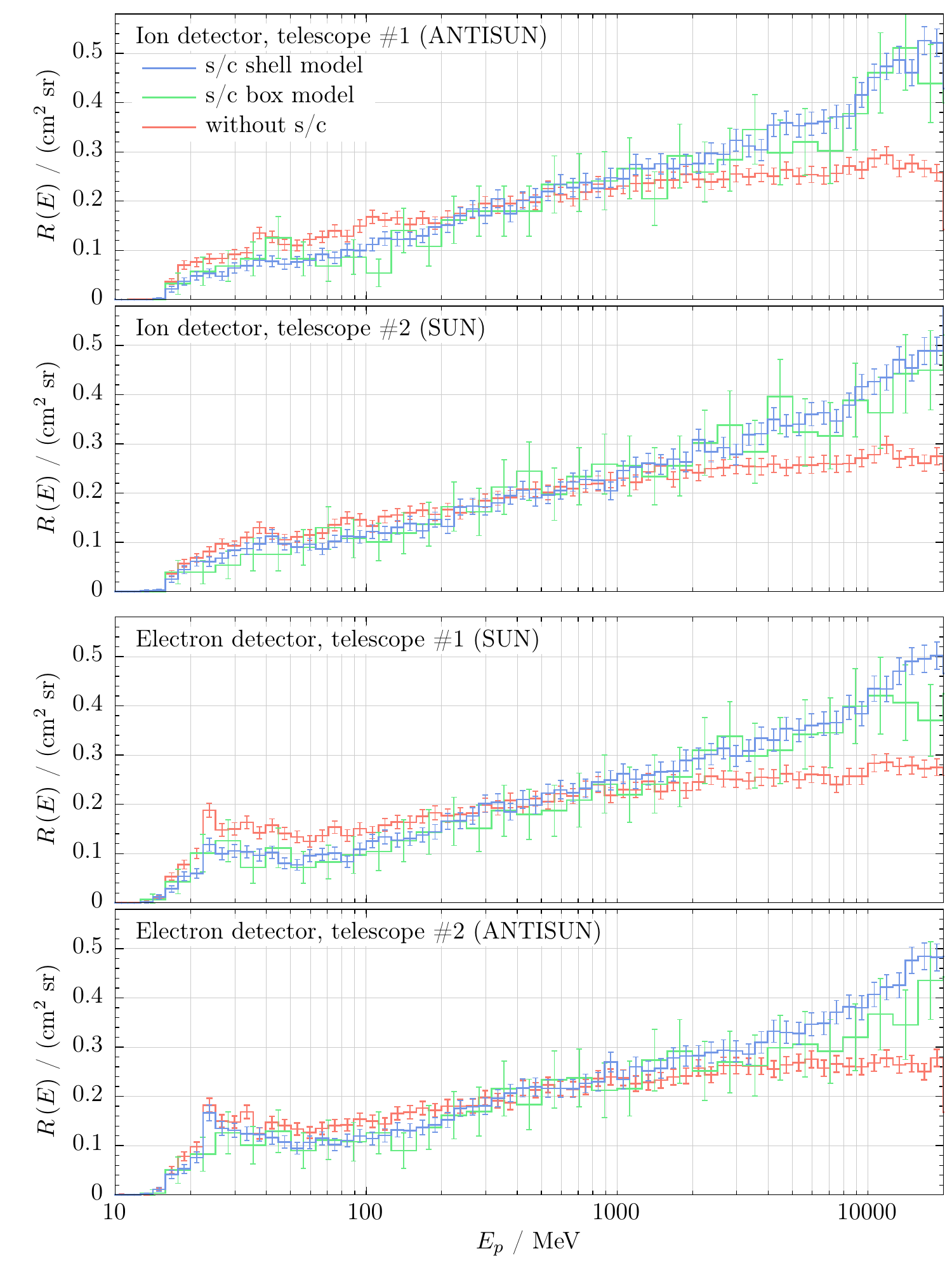}
\caption{Comparison of the computed response for particles entering the detector system from the side (path \protect\circled{2} in Fig. \ref{fig:SSD}) for three different kind of SEPT simulation models: the shell models for telescope \#1 and \#2, the box model, and the model of the SEPT without any s/c influence. The response is plotted against the primary energy $E_p$ of the protons. The top two panels show the result for the ion detectors of telescope \#1 and \#2. The lower two panels show the result for the electron detectors.}
\label{fig:run-comparison}
\end{figure}
Figure \ref{fig:run-comparison} displays from top to bottom the responses for the ion and electron detectors of both telescopes of SEPT-E on STA for protons entering the detector system from the side (see Fig. \ref{fig:sketch}). The red, green and blue histograms show the results when neglecting the s/c, utilizing the box- and the shell model, respectively. The uncertainty of the responses were calculated using the statistical error $\sqrt{N_i}$ of the simulation, where $N_i$ are the number of counted particles of the $i$-th energy bin on the $E_p$ axis during the simulation. Below $\sim200$~MeV the shielding effect of the s/c is clearly visible by a reduced response for the box model and shell models compared to the basic instrument model.
However, at energies above $\sim$2~GeV the production of secondaries in the s/c becomes significant leading to an increase of the response when utilizing both the box- and shell model with respect to the one that does not include the s/c. Taking into account the agreement between both s/c models as well as the uncertainty of the mass distribution within the spacecraft, we conclude that the shell model sufficiently reproduces the influence of the s/c. The differences at higher energies can be neglected taking into account that the energy spectra of SEPs are typically soft at energies above a few hundreds of MeV and that at energies above 4.5~GeV the Galactic Cosmic Ray (GCR) spectrum falls with $E^{-2.7}$, resulting in  minor contributions to the count rates.

\subsection{SEPT Response Functions for Protons}
Utilizing the improved SEPT-E shell models described above we computed the response functions of the two proton and electron detectors for isotropic incident of protons from 20~keV to 20~GeV. In what follows we focus our analysis on the proton detector of telescope \#2, which reflects to the SUN direction. To distinguish the result for the different particle tracks sketched in Fig. \ref{fig:SSD} the counted hits of the simulation were categorized by three categories of the primary particle direction: \circled{1} magnet, \circled{2} side, and  \circled{3} foil as sketched in Fig. \ref{fig:sketch}.

\begin{figure}
\centering
\includegraphics[width=\cw]{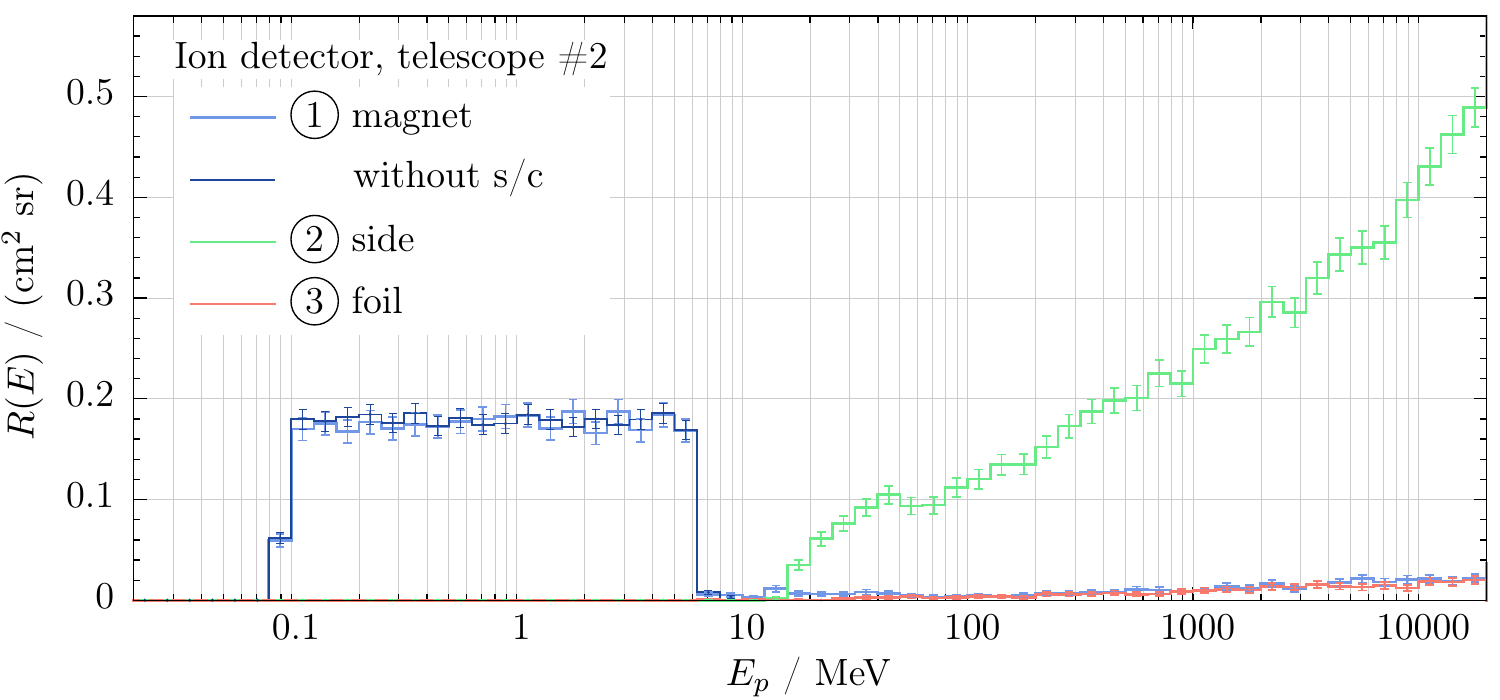}
\caption{Total response $R(E)$ of the ion detector of telescope \#2 of SEPT-E on STA for protons in the energy range from 20~keV to 100~MeV computed utilizing the shell model. Previous simulation results for the magnet direction using the model that neglects the s/c are shown in a darker blue as well.}
\label{fig:sept-e-sta-nominal-geo-log}
\end{figure}

The light blue line in Fig. \ref{fig:sept-e-sta-nominal-geo-log} shows the total response of the SEPT-E telescope \#2 ion detector for the nominal direction \circled{1} (``magnet"). If neither a gap between both detectors nor the crosstalk ring existed, this response function would be the one of an ideal telescope. For the nominal direction, which is unaffected by the adaptations made to the model, the results are in agreement with those from previous simulations using the simplified instrument model from \citet{Mueller-Mellin-etal-2008}. Those previous results are displayed in Fig.~\ref{fig:sept-e-sta-nominal-geo-log} with the dark blue lines. The red curve in the figure shows the response $R(E)$ for protons passing through the electron SSD from the opposite side. This contribution is small because protons need to cross the 100~$\mu$m wide crosstalk ring on the electron SSD in order to avoid the anticoincidence trigger. These particles represent tracks \circled{3} in  Fig. \ref{fig:SSD}. Note that only protons with energies above a few MeV contribute to the count rate. At these energies a similar contribution can be seen in the magnet category as well. These particles cross the crosstalk ring after they penetrated the ion detector. As discussed in the previous section, protons hitting the instrument from the side can trigger the center elements without triggering the guard ring (see track \circled{2} in Fig.~\ref{fig:SSD}). The contribution of this category has been discussed in the previous section and is shown by the  green line in Fig.~\ref{fig:sept-e-sta-nominal-geo-log}.

\begin{figure}
\centering
\includegraphics[width=\cw]{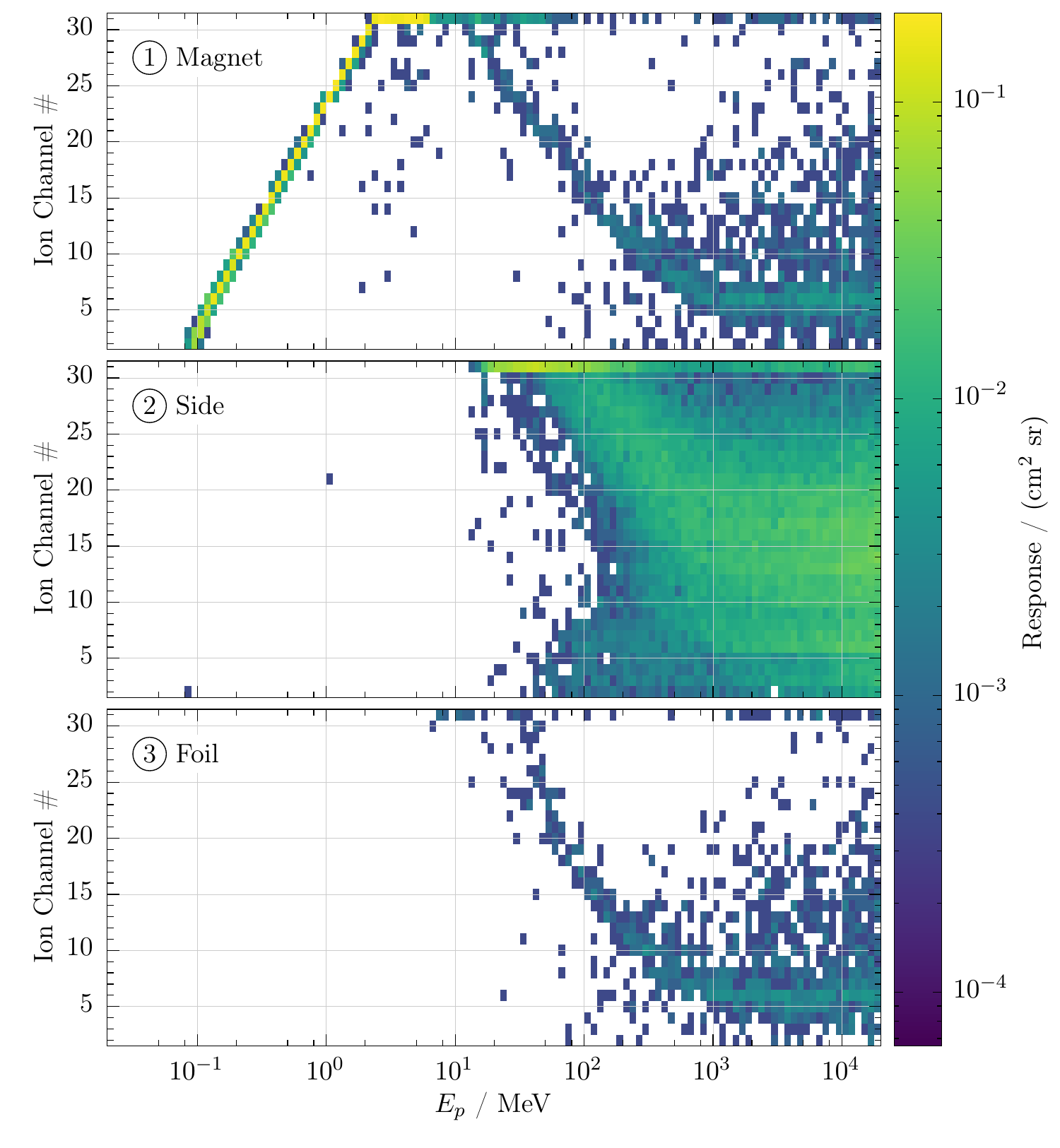}
\caption{Proton response matrix of the SEPT-E on STA telescope \#2 ion channels separated by the the particle's angle to the detector axis into the three categories magnet \protect\circled{1}, side \protect\circled{2}, and foil \protect\circled{3} (also see Figures \ref{fig:sketch} and \ref{fig:SSD}).}
\label{fig:sept-e-sta-response-matrix}
\end{figure}

While Fig.~\ref{fig:sept-e-sta-nominal-geo-log} shows the total response, the individual responses of the ion channels are plotted together as the response matrix shown in Fig.~\ref{fig:sept-e-sta-response-matrix}. This figure displays $R(E)$ of each channel as a function of the primary proton kinetic energy from 20~keV to 20~GeV on the x-axis and with the proton channel number from \#2 to \#31 on the y-axis. Values from below $10^{-4}$\cmcmsr{} to 0.2\cmcmsr{} are color coded on a logarithmic scale from blue to yellow. Again, the responses were categorized into the directions \circled{1} magnet, \circled{2} side and \circled{3} foil. As expected the instrument measures low energy protons (below 10~MeV) from the nominal direction only. However, major contributions from the side are expected for all ion channels. Potential contributions come from Galactic Cosmic Rays (GCR) and ``strong" SEP events.

Using the instrument response, count rates $C_i$ of $m$ channels $i=\{1, ..., m\}$ for an arbitrary isotropic particle input distribution $J(E)$ are computed by the following equation:
\begin{equation}
C_i = \sum\limits_{n}\int\limits_0^\infty \mathrm{d}E \, (R_{n,i}(E) \, J_n(E))\,,
\label{eq:countrates}
\end{equation}
with $R_{n,i}(E)$ the energy response of the $i$-th channel to a particle of type $n$ and $J_n(E)$ the a priori unknown energy spectra of that type. Here we only investigate the influence of protons, assuming that the impact of electrons and $\alpha$-particles can be neglected. Note that during quiet times these assumptions (major contributions by protons and isotropy) are reasonably well fulfilled. In order to estimate the goodness of our simulations we compute the quiet time count rate variation from the beginning of 2007 to the end of 2009 and compare our results to corresponding measurements.

It is important to note that the statistics of the simulations using the box model, as shown in the previous section in Fig. \ref{fig:run-comparison}, are not sufficient to derive response matrices with an acceptable resolution and uncertainty for this analysis.

\section{Quiet Time Measurements} \label{sec:quiet-time-measurements}
The unusual long solar minimum from 2007 to 2010 \citep[see for example][and references therein]{Mewaldt-etal-2010, Oh-etal-2013} allows not only to study the variation of GCRs but also the investigation of the origin of suprathermal ions \citep{Gloeckler-etal-2008, Murphy-etal-2017}. As mentioned above ion measurements from SEPT are contaminated by  protons with energies above $\sim100$~MeV (see Fig.~\ref{fig:sept-e-sta-response-matrix}). During solar minimum periods the major source of these ions are galactic cosmic rays. Their energy flux spectra can be estimated by the so called force field solution (FFS) (see Eq.~\ref{eq:force-field-equation}). This equation was derived by \citet{Gleeson-Axford-1968} and others in order to describe the transport of GCR ions in the heliosphere: 

Following \citet[]{Moraal-2013}(see also \citet{Gleeson-Axford-1968, Caballero-Lopez-Moraal-2004}), the transport of GCRs in the heliosphere can be approximated by a simple 1-dimensional convection-diffusion equation for the phase space density $f$: 
\begin{equation}
\label{eq:conv-diff}
\frac{v \, P}{3}\frac{\partial f}{\partial P} + \kappa \frac{\partial f}{\partial r} = 0\,, 
\end{equation}
with $v$, $r$, $P$ denoting the solar wind speed, the radial distance and the particle rigidity, respectively. If the diffusion coefficient $\kappa(r,P)$ is separable $\kappa=\kappa_1(r)\cdot \kappa_2(P)$ with $r$ the heliocentric distance and $P$ the particle rigidity, and furthermore $\kappa_2(P) \propto P$, an analytical solution for Eq.~\ref{eq:conv-diff} exists for $J(E)=\frac{f}{v}$: 
\begin{equation}
J(E,\phi) = J_{\mathrm{LIS}}(E+\Phi)\frac{(E)(E+2\,E_r)}{(E+\Phi)(E+\Phi+2 \,E_r)}
\label{eq:force-field-equation}
\end{equation}
The force field function $\Phi$ is given by $\Phi = (Z e/A)\phi$, where $Z$ and $A$ are the charge and mass number of the cosmic ray nuclei, respectively, leaving the modulation parameter $\phi$ as the only temporal variable. $E$ represents the kinetic energy of the particles, $E_r$ their rest energy ($E_r=938$~MeV for protons) and $J_{\mathrm{LIS}}(E)$ gives the differential energy spectra of the LIS representing the boundary condition of the force field approximation. In this work we utilize the LIS originally proposed by \citet{Burger-et-al-2000} and also described by \citet{Usoskin-etal-2005} and the new two parameter ($\phi_{pp}$, $\phi_{\mathrm{Uso11}}$) model introduced by \citet{Gieseler-etal-2017}. 
Each pair ($\phi_{pp}$, $\phi_{\mathrm{Uso11}}$) is used to compute a  rigidity-dependent modulation parameter $\phi(P)$ as described by these authors. During the unusual long solar minimum the modulation potential varied for ions below 3~GV and above 10~GV from $\phi_{pp}=0.54$~GV and $\phi_{\mathrm{Uso11}}=0.39$~GV in January 2007 to $\phi_{pp}=0.4$~GV and $\phi_{\mathrm{Uso11}}=0.25$~GV in December 2009, respectively.

\begin{figure}
\centering
\includegraphics[width=\cw]{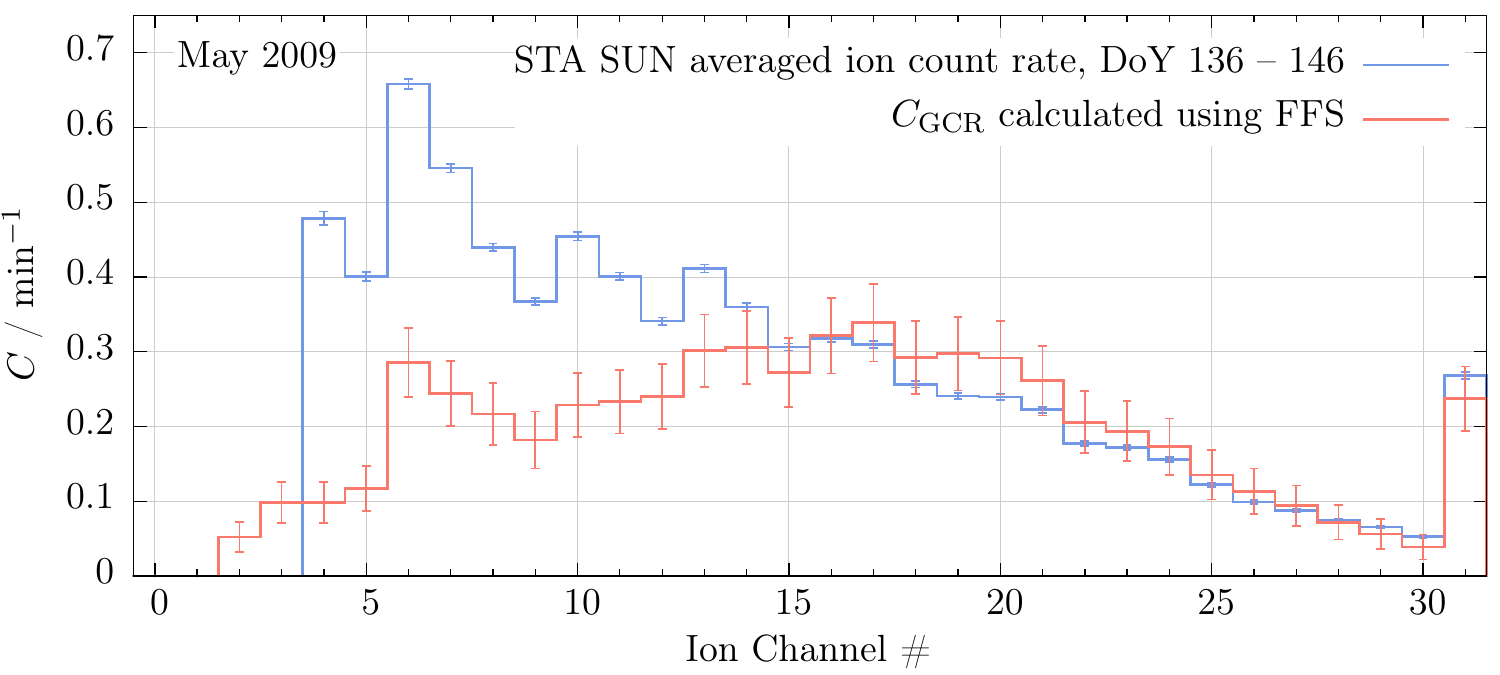}
\caption{Red line: calculated count rate for the ion channels using the corresponding response matrix (Fig. \ref{fig:sept-e-sta-response-matrix}) and a GCR force-field-solution spectrum utilizing the rigidity dependent modulation parameter $\Phi(P)$ for May 2009 from \citet{Gieseler-etal-2017}. Blue line: Measured quiet time count rate for the ion channels of SEPT SUN on STA for May 2009, derived as an average value from DoY 136 to 146.}
\label{fig:gcr-counts}
\end{figure}

Utilizing the FFS (Eq. (\ref{eq:force-field-equation})) with the rigidity dependent modulation parameter $\phi(P)$ from \citet{Gieseler-etal-2017} together with SEPT's response matrix (Fig. \ref{fig:sept-e-sta-response-matrix}), we computed the predicted count rates $C_{\mathrm{GCR}}$ induced by GCR on SEPT for the 32 ion channels using Eq. (\ref{eq:countrates}). Figure \ref{fig:gcr-counts} displays $C_{\mathrm{GCR}}$ for May 2009 by the red histogram. The errors were calculated from the statistical error of the simulation only, hence from the uncertainty of the calculated instrument responses. To compare our computations with measured data, quiet time periods had to be determined. In order to select quiet times we used the following method:

\begin{figure}
\centering
\includegraphics[width=\cw]{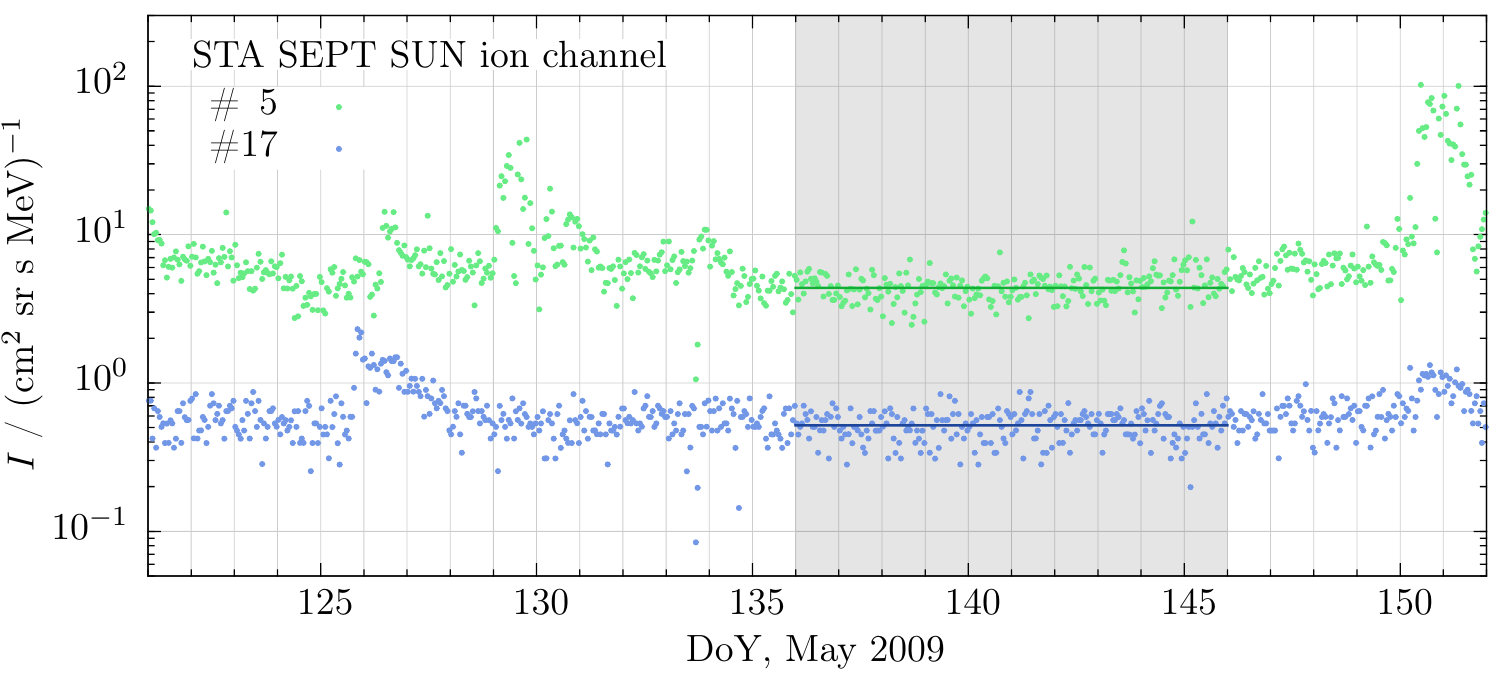}
\caption{60~min-averaged intensities of the ion channels \#5 (110--118.6~keV) and \#17 (438.1--496.1~keV) measured by SEPT SUN on STA during May 2009. The time frame from DoY 136 to 146, as marked in gray, was selected to calculate an average quiet time count rate for this month. The average values are shown by the darker straight lines.}
\label{fig:gcr-2009-5-selection}
\end{figure}

Figure~\ref{fig:gcr-2009-5-selection} shows the flux time profiles of 114~keV and 466~keV protons (ion channels \#5 and \#17, respectively) as measured by SEPT SUN aboard STA during May 2009. We define a day of data as a day of quiet time whenever more than 90\% of the 24 60-minute-averaged data points per day of these ion channels \#5 and \#17 are concurrently below 7\intunit{} and 0.8\intunit, respectively. These two limits have been determined by analyzing count rate histograms for this solar minimum. For each month we then selected the longest interval of coherent quiet days and discarded the first and last day of each interval. Further we only took into account periods with at least three coherent quiet days. In Fig.~\ref{fig:gcr-2009-5-selection} the interval selected as quiet time by this method is marked in gray and ranges from DoY 136 to 146. The average over this interval was used to derive the quiet time count rate for the month. The blue line in Fig.~\ref{fig:gcr-counts} shows those mean values. The displayed error bars of the measured data are the calculated standard errors of the corresponding mean value. Comparing the measured and calculated count rates in Fig.~\ref{fig:gcr-counts} it becomes evident that the count rate of all channels above \#16 is dominated by the GCR contribution. We note that the predicted count rates $C_\mathrm{GCR}$ are mostly given by protons entering the detector system on oblique tracks from the side (see Fig. \ref{fig:sketch} and \circled{2} in Fig. \ref{fig:SSD} and \ref{fig:sept-e-sta-response-matrix}) which confirms the necessity of the new simulation model for this analysis.

\subsection{GCR Modulation}
\begin{table}
\caption{Quiet time intervals for all months of 2007, 2008, and 2009 determined by the method described in the text (Sec. \ref{sec:quiet-time-measurements}) for SEPT SUN on STA. For empty/missing months it was not possible to determine a quiet time interval with a length of at least three coherent days.}
\begin{tabular}{rcr}
\hline
\multirow{2}{*}{Month} & Determined quiet time interval & \multirow{2}{*}{\# of days} \\
 & (as fractional DoY) & \\
\hline
Jun--2007 & 167.0 -- 170.0 &  3 \\
Jul--2007 & 188.0 -- 191.0 &  3 \\
Aug--2007 & 230.0 -- 236.0 &  6 \\
Sep--2007 & 252.0 -- 255.0 &  3 \\
Oct--2007 & 277.0 -- 282.0 &  5 \\
Nov--2007 & 305.0 -- 313.0 &  8 \\
Dec--2007 & 357.0 -- 366.0 &  9 \\ & & \\
Jan--2008 &  18.0 -- 32.0  &  14\\
Feb--2008 &  52.0 -- 55.0  &  3 \\
Mar--2008 &  84.0 -- 87.0  &  3 \\
Apr--2008 & 109.0 -- 114.0 &  5 \\
May--2008 & 147.0 -- 150.0 &  3 \\
Jun--2008 & 173.0 -- 177.0 &  4 \\
Jul--2008 & 210.0 -- 214.0 &  4 \\
Aug--2008 &       -        &  - \\
Sep--2008 &       -        &  - \\
Oct--2008 & 292.0 -- 301.0 &  9 \\
Nov--2008 &       -        &  - \\
Dec--2008 & 359.0 -- 367.0 &  8 \\ & & \\
Jan--2009 &  ~8.0 -- 13.0  &  5 \\
Feb--2009 &  35.0 -- 47.0  &  12\\
Mar--2009 &  60.0 -- 67.0  &  7 \\
Apr--2009 &  91.0 -- 96.0  &  5 \\
May--2009 & 136.0 -- 146.0 &  10\\
Jun--2009 & 166.0 -- 170.0 &  4 \\
Jul--2009 & 185.0 -- 193.0 &  8 \\ 
Aug--2009 & 213.0 -- 222.0 &  9 \\
Sep--2009  & 262.0 -- 268.0 &  6 \\
Oct--2009 & 297.0 -- 300.0 &  3 \\
\hline
\end{tabular}
\label{tab:quiet-time-intervals}
\end{table}

\begin{figure}
\centering
\includegraphics[width=\cw]{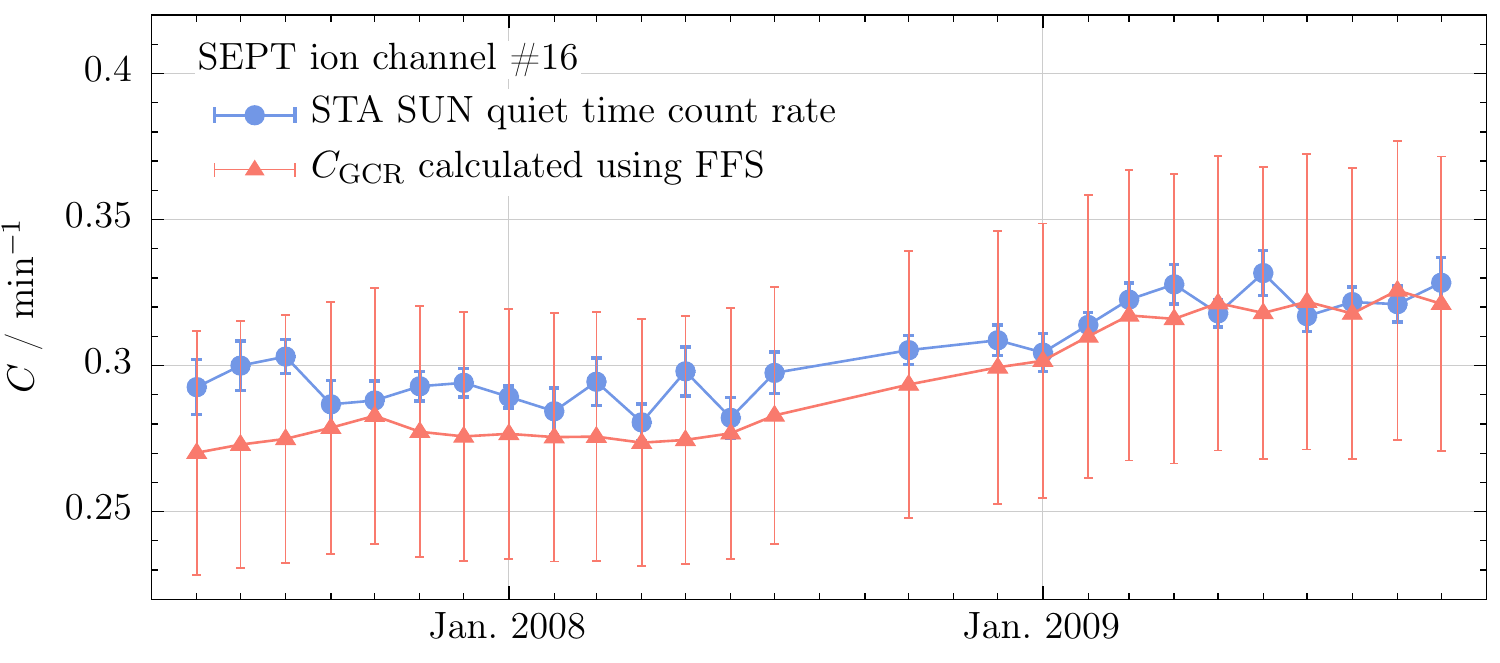}
\caption{Development of the uncorrected (blue line) quiet time count rate of ion channel \#16 ($\sim400$~keV) during the solar minimum from June 2007 to October 2009. The predicted count rate $C_\mathrm{GCR}$ is shown in red.}
\label{fig:gcr-quiet-times}
\end{figure}

This method has been applied to determine all quiet times during the solar minimum from 2007 to the end of 2009 for SEPT on STA. The resulting quiet time intervals and their lengths are summarized in Tab. \ref{tab:quiet-time-intervals}. We note that for some months it was not possible to determine quiet times because of continuous flux enhancements. These quiet time intervals are utilized to compare the calculated GCR induced count rates with the observed ones. Figure \ref{fig:gcr-quiet-times} shows the development of such quiet time count rate for the ion channel \#16 ($\sim400$~keV) of SEPT SUN on STA from June 2007 to October 2009 (blue) together with the prediction (red) utilizing the FFS and the modulation potentials from \citet{Gieseler-etal-2017}. From Figs.~\ref{fig:gcr-counts} and \ref{fig:gcr-quiet-times} it is evident that the SEPT energy spectrum for all channels above \#16 is determined by the imprint of GCRs.

\subsection{Quiet Time Spectrum for 2007}
\begin{figure}
\centering
\includegraphics[width=\cw]{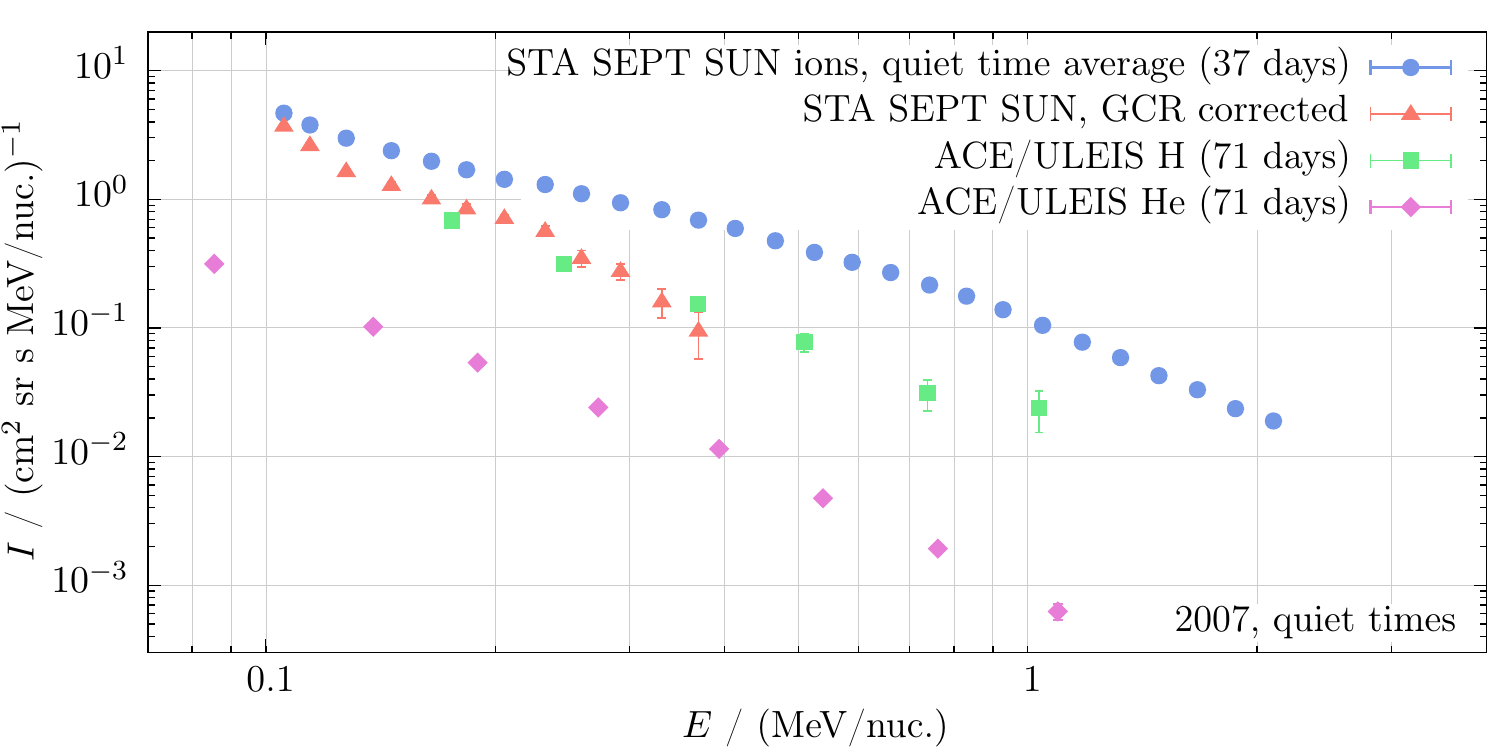}
\caption{Uncorrected (blue points) and GCR corrected (red points) quiet time spectra from SEPT SUN on STA for 2007 calculated as an average of the determined 37 quiet days of 2007 listed in Tab. \ref{tab:quiet-time-intervals}. For comparison, similar spectra for protons and helium from 71 days of 2007 measured by the ACE/ULEIS instrument are shown as well (taken from \citet{Mason-2012}).}
\label{fig:2007-corrected-spectrum}
\end{figure}

From Fig. \ref{fig:gcr-counts} it is evident that the contribution of GCRs decreases with decreasing channel number for all channels below \#16. Figure \ref{fig:2007-corrected-spectrum} displays the quiet time spectrum measured by SEPT SUN aboard STA during quiet time periods with (red symbols) and without (blue symbols) correction for the GCR contribution. The corrected spectra was determined by subtracting the GCR induced count rates $C_\mathrm{GCR}$, calculated as described above and averaged from June to December 2007, from the uncorrected data. As expected the corrections are small at low energies and reach an order of magnitude for ions above 300~keV. \citet{Mason-2012} published proton and helium spectra during quiet times in 2007 measured by the Ultra Low Energy Isotope Spectrometer \citep[ULEIS,][]{Mason-etal-1998} on the Advanced Composition Explorer \citep[ACE,][]{Stone-etal-1998} spacecraft. The corresponding fluxes for protons and helium are shown by the green and purple symbols, respectively. Taking into account the uncertainties, both ULEIS and SEPT are in very good agreement in the energy range from about 150 to 250~keV \citep[for details see][and references therein]{Wraase-etal-2018}. Though helium is counted in the SEPT ion channels too, it  has a four times higher energy loss than protons at the same energy per nucleon. Therefore a notable contribution from primary helium to the SEPT measurements is only expected for bins larger than number \#26 corresponding to energy losses above 1.3~MeV. At lower energy losses only secondary particles i.e. secondary protons should contribute too. However, the flux of these secondary particles is expected to be negligible in comparison to the one of the primary protons \citep[see also][]{Kuehl-etal-2015}.

\section{Summary and Conclusion} \label{S-Conclusion} 
The response function for protons impinging on the four telescopes of the SEPT have been calculated. A simplified model of the s/c shielding by using a spherical shell has been developed and validated against a more complex s/c model approximation. In order to ascribe the measurements during quiet time periods the contributions of GCR protons are computed utilizing the proton intensities at 1 AU determined by the model of \citet{Gieseler-etal-2017}. We found that for all SEPT energy channels above $\sim$400~keV (\#16) the quiet time flux is fully described by the induced GCR signal. At lower energies we could show that the ion spectrum is in very good agreement with the one published by \citet{Mason-2012}.  

\begin{acks}
The STEREO/SEPT project is supported under Grant 50~OC~1302 by the German Bundesministerium f\"ur Wirtschaft through the Deutsches Zentrum f\"ur Luft- und Raumfahrt (DLR). P.K. has received funding from the European Union's Horizon 2020 research and innovation programme under grant agreement No~637324.
\end{acks}

\bibliographystyle{spr-mp-sola}
\bibliography{sept}  

\IfFileExists{\jobname.bbl}{} 
{
\typeout{}
\typeout{****************************************************}
\typeout{****************************************************}
\typeout{** Please run "bibtex \jobname" to obtain}
\typeout{** the bibliography and then re-run LaTeX}
\typeout{** twice to fix the references !}
\typeout{****************************************************}
\typeout{****************************************************}
\typeout{}
}

\end{article} \end{document}